\documentclass{iaus}
\usepackage{graphicx}
\usepackage{iauvoltoc}
\title[Organic matter in disks] 
{The birth and death of organic molecules in \\ protoplanetary disks}

\author[Th.~Henning and D.~Semenov]   
{Thomas Henning \and Dmitry Semenov}

\affiliation{Max Planck Institute for Astronomy \\
Koenigstuhl 17, D-69117 Heidelberg, Germany \\ email: {\tt henning,semenov@mpia.de}
}

\pubyear{2008}
\volume{251}  
\jname{Organic Matter in Space}
\editors{S. Kwok \& S. Sandford, eds.}
\begin{document}

\maketitle

\aindex{Henning, T.}
\aindex{Semenov, D.}
\addcontentsline{toc}{titl}{The birth and death of organic molecules in protoplanetary disks}
\addcontentsline{toc}{auth}{T.~Henning \and D.~Semenov}

\begin{abstract}
The most intriguing question related to the chemical evolution of protoplanetary disks is the genesis of pre-biotic organic molecules in the planet-forming zone. In this contribution we briefly review current observational knowledge of physical structure and chemical composition of disks and discuss whether organic molecules can be present in large amounts at the verge of planet formation. We predict that some molecules, including CO-bearing species such as H$_2$CO, can be underabundant in inner regions of accreting protoplanetary disks around low-mass stars due to the high-energy stellar radiation and chemical processing on dust grain surfaces. These theoretical predictions are further compared with high-resolution observational data and the limitations of current models are discussed.
\sindex{astrochemistry}
\sindex{line: formation}
\sindex{molecular data}
\sindex{molecular processes}
\sindex{radiative transfer}
\sindex{(stars:) circumstellar matter}
\sindex{(stars:) planetary systems: protoplanetary disks}
\sindex{stars: pre--main-sequence}
\sindex{submillimeter}
\sindex{X-rays: stars}
\end{abstract}

\section{Introduction}
The origin and evolution of life as we know it are tightly related to the chemistry of complex carbon-bearing molecules. While the transition from macromolecules to the simplest living organisms has likely proceeded on Earth, we do not know yet organic molecules of what complexity have been available during build-up phase of the primordial/secondary Earth atmosphere and oceans. During the last few decades a multitude of species, including alcohols (e.g. CH$_3$OH), ethers (e.g. CH$_3$OCH$_3$), acids (e.g. HCOOH) have been discovered in interstellar space, with atomic masses up to a few hundred\footnote{\it{http://astrochemistry.net/}} (for a recent review see \cite{Snyder_06}). A precursor of amino acids, amino acetonitrile, and the simplest sugar, glycolaldehyde, have been found toward the star-forming region Sagittarius B2(N) (\cite{Hollis_ea04,Belloche_ea08}). Thus, many simple ``blocks'' of prebiotic molecules do exist in space. It is natural to ask what happens to these species during the prestellar/hot core phase, passage of an accretion shock, and inside a protoplanetary disk. Are organic molecules present in large amounts in circumstellar disks at the verge of planet formation? Could they form and survive in such a harsh environment as an accretion disk?

Despite the variety of ``interstellar'' molecules, only formaldehyde (H$_2$CO) and a few other non-organic species have been detected and spatially resolved with interferometers in several nearby protoplanetary disks (e.g., \cite{DGG97,Kastner_ea97,Aikawa_ea03,Qi_ea03,Dutrey_ea07}). These multi-molecule, multi-transition studies allowed to constrain basic disk parameters like radii, masses, kinematics, temperature and density profiles, ionization degree and depletion factors (e.g., \cite[Dartois et~al.~2003]{DDG03},\cite{Semenov_ea05},\cite[Qi et~al.~2006]{Qi_ea06},\cite[Pi\'etu et~al.~2007]{PDG_07},\cite{Qi_ea08}). Using advanced chemical models and indirect observational evidence, one can get a clue about the presence of other, yet undetected organic molecules in disks and estimate their abundances.

It is commonly believed now that the formation of complex (organic) molecules starts already in cold dense cloud cores on dust grain surfaces serving as catalysts for many exothermic reactions between radicals and light atoms, with formaldehyde being one of the precursors for complex organic molecules. The newly produced species can be eventually returned into the gas phase either during the slow heat-up phase after the formation of a central star (\cite[Garrod \& Herbst 2006]{Garrod_ea06}) or due to some thermal/non-thermal desorption mechanisms, like cosmic rays/X-rays/UV heating of grains (\cite{dHendecourtea82,Leger_ea85,ShalabeiaGreenberg94,Najitaea01,Garrod_ea07}).

Transformation of a cloud into an actively accreting disk caused by gravitational collapse and angular momentum transport further modifies the composition of gas and dust as it passes through a shock front (\cite{Lada_85,Hassel_04}). Furthermore, dynamical transport in accretion disks can also be efficient for enriching the gas with complex species through evaporation of icy mantles in warmer, less opaque regions (\cite{Willacy_ea06,Semenov_ea06}). The UV radiation from the star and interstellar radiation field plays a major role in disk chemistry by dissociating and ionizing molecules, and heating the gas above the midplane where many molecular lines get excited (e.g., \cite{vZea03}). Disk chemistry by itself can lead to the production of complex organic molecules.

The recent detection of Ne{\sc II} line emission from several protoplanetary disks by {\it Spitzer} (\cite{Pascucci_ea07}) supports theoretical predictions of \cite{Glassgold_ea07} that the upper disk parts can be ionized and heated by the intense X-ray radiation from a young star. This thermal bremsstrahlung radiation is likely produced in reconnection loops in stellar corona at a distance of up to 0.1~AU from the star and is capable of penetrating deeply into the disk inner region -- the zone where planets form (e.g., \cite{IG99}). This X-ray radiation ionizes helium atoms, which destroy CO and replenish thedisk gas with  ionized atomic carbon. This leads to the formation of heavy cyanopolyynes and long carbon chains, partly on grain surfaces, which may lock a significant fraction of elemental carbon in the inner disk region (\cite[Semenov et~al.~2004]{Red2}). Thus it is of utter importance to reveal which mechanisms and processes are important during various stages of protoplanetary disk evolution, using advanced theoretical modeling and high-quality observational data.

In the framework of the ``Chemistry In Disks'' (CID) collaboration between groups in Heidelberg, Bordeaux, Paris and Jena, we have initiated a program to study and characterize chemical evolution and physical properties of nearby protoplanetary disks surrounding young stars of various masses and ages (see, e.g., \cite{Dutrey_ea07}). Among the various goals of the project, we searched for emission lines of precursors for complex organic molecules and detected and resolved the disk around the low-mass star DM Tau in the H$_2$CO(3-2) line with the Plateau de Bure interferometer (though with modest SNR of $\sim 3-5$). We found that the H$_2$CO emission is not centrally peaked as is the dust continuum, but shows an asymmetric, ring-like structure with a large inner ``hole'' of $\sim 100$~AU. The high-resolution observations of DM Tau with the IRAM interferometer by \cite{DDG03} and \cite{PDG_07} did not reveal the presence of central depression either in CO lines or in dust continuum. In this contribution we review our current knowledge about chemistry in disks and predict that the H$_2$CO inner hole, if it really exists, should likely be caused by \emph{chemical effects}, which we will discuss in the paper.

\section{Observational facts}
Up to now more than 150 molecules have been discovered in space\footnote{\it{http://astrochemistry.net/}}. Among them only a small fraction have been detected in planet-forming disks with the aid of millimeter interferometry: CO and its isotopes, CN, HCN, DCN, HNC, H$_2$CO, C$_2$H, CS, HCO$^+$, H$^{13}$CO$^+$, DCO$^+$, and N$_2$H$^+$ (\cite{DGG97,Kastner_ea97,Aikawa_ea03,DGH_07,Qi_ea08}). A few bright and large disks like surrounding DM Tau, LkCa~15, and MWC~480 have been investigated in detail in a dozen of molecular transition and dust continuum. Combined analysis of these line and continuum data allowed us to derive disk sizes, orientation, kinematics, distribution of temperature, surface density, and molecular column densities (e.g., \cite[Dartois et~al.~2003]{DDG03},\cite[Qi et~al.~2006]{Qi_ea06},\cite{Isella_ea07},\cite[Pi\'etu et~al.~2007]{PDG_07}).

The lines of the abundant CO molecule serve as a probe of disk geometry as well as thermal structure and surface density distribution and kinematics. Due to selective photodissociation, disks appear increasingly larger in the dust continuum and the C$^{18}$O, $^{13}$CO, and $^{12}$CO lines, with a typical radius of $\sim$ 300--1\,000~AU (\cite{DGH_07}). An important finding is the presence of vertical temperature gradients in many disks, as predicted by physical models, while a few disks with large inner cavities do not show evidence for such a gradient (e.g., GM~Aur). Furthermore, disks around hotter Herbig Ae/Be stars are systematically warmer than those around cool Sun-like T Tau stars. Recently, \cite{Qi_ea06} have found that the observed intensity ratios of the CO low- to high-level lines in the TW~Hya disks require an additional heating source, which could be the X-ray stellar radiation.

The observed lines of C$_2$H, CN, and HCN are sensitive to the properties of the impinging UV radiation, in particular to the fraction of the total UV luminosity emitted in the Ly$_\alpha$ line (\cite{BCDH03}). The observed ratio of the CN to HCN column densities in disks is typical of photon-dominated chemistry, as predicted by the chemical models (Chapillon et al. 2008, in prep.). Molecular ions (HCO$^+$ and N$_2$H$^+$) are the dominant charge carriers at intermediate disk heights and their observations allowed to constrain the ionization fraction in these regions, with a typical value of $10^{-8}$ (\cite{Qi_ea03,Dutrey_ea07}). The observed ratios of DCO$^+$ to HCO$^+$ and DCN to HCN column densities have much higher D/H value than the cosmic abundance of $\sim 0.01\%$ and thus deuterium fractionation is effective in disks (\cite{Qi_ea08}).

In general, the observed molecular abundances are lower by factors 5--100 in protoplanetary disks compared to the values in Taurus molecular cloud, likely due to efficient freeze-out and photodissociation. A puzzling observational fact is that a significant reservoir of very cold CO and HCO$^+$ gas exists in the disks of DM~Tau and LkCa~15, at temperatures $\lesssim$ 13-17~K, which are below the freeze-out temperature of CO (20~K). Conventional chemical models cannot explain this fact without invoking a non-thermal desorption mechanism that works in the dark disk midplane, like efficient turbulent diffusion and UV-photodesorption driven by cosmic ray particles (\cite{Semenov_ea06,Oeberg_ea07}).

Observations of dust thermal emission at (sub-) millimeter and centimeter wavelengths are used to measure the slope of the wavelength dependence of the dust opacities that is a sensitive indicator of grain growth and sedimentation in disks (e.g., \cite{Rodmann_ea06}). There is strong evidence that in many evolved disks, with ages of a few Myrs, dust grains grow until at least pebble-like sizes. The results from the {\it Infrared Space Observatory} and {\it Spitzer} reveal the presence of a significant amount of frozen material and a rich variety of amorphous and crystalline silicates and PAHs in disks (e.g., \cite{vdAea00,vanDishoeck_ARA2004,Bouwman_ea08}). The PAH emission features at near- and mid-infrared wavelengths are excited by the incident stellar radiation field and as such depend on disk vertical structure and turbulent state (\cite{Dullemond_ea07}). These lines are more easily observed in disks around hot, intermediate-mass Herbig Ae/Be stars as compared to cool, Sun-like T Tauri stars (e.g., \cite{Acke_ea04,Geers_ea07,Sicilia-Aguilar_ea07}).

Various solid-state bands observed at 10--30$\mu$m in emission belong to amorphous and crystalline silicates at $T\gtrsim 100-300$~K with varying Fe/Mg ratios and grain topology/sizes (e.g., \cite{vanBoekel_ea04,Natta_ea07,Bouwman_ea08,VH_08}). The composition of the hot gas in the inner disk as traced by ro-vibrational emission lines from CO, CO$_2$, C$_2$H$_2$, HCN and recently H$_2$O and OH, suggests that complex chemistry driven by endothermic reactions is at work there (\cite{Bea03,Lahuis_ea06,Eisner_ea07,Salyk_ea08}). At larger distances from the star the disk becomes colder and most of these molecules stick to dust grain, forming icy mantles. The main mantle component is water ice with trace amount of other more volatile materials like CO, CO$_2$, NH$_3$, CH$_4$, H$_2$CO, HCOOH and CH$_3$OH (\cite{Zasowski_ea08}). Typical relative abundances of these minor constituents are about 0.5--10\% of that of water.

\section{Chemical structure of a disk}
\begin{figure}
\begin{center}
 \includegraphics[width=2.5in,angle=270]{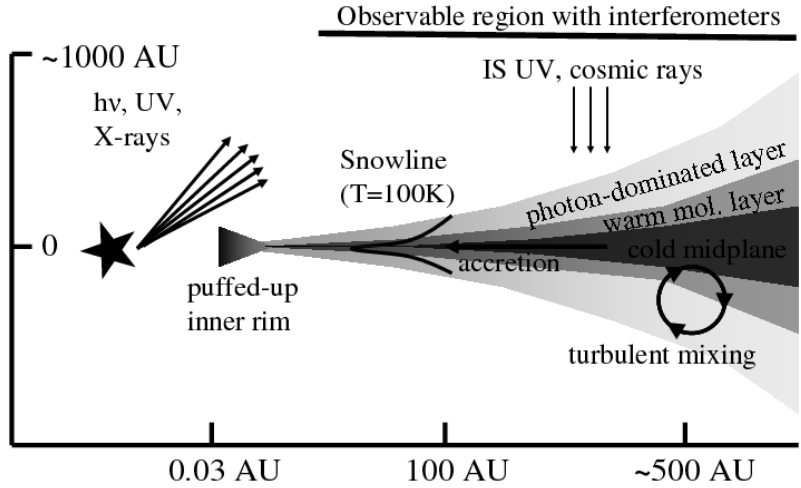}
\caption{Physical and chemical structure of a protoplanetary disk. In the dark, dense and cold midplane most of molecules reside on dust grains, and chemical evolution is dominated by ion-molecule and surface reactions. This region has the lowest ionization degree, with dust grains being the most abundant charged species. A warmer intermediate layer is located above midplane. It is heated by mild UV radiation. Many reactions with barriers can occur and a rich variety of molecules exist in the gas phase. This is the zone where most of the molecular lines are formed. The ionization fraction in the intermediate layer is determined by a multitude of molecular ions, in particular HCO$^+$. Further above, a hot, highly ionized disk atmosphere is located, where only the simplest radicals and ions (apart from H$_2$) survive. This is the region where ionized carbon is abundant and C$^+$ emission lines are excited.}
   \label{scheme}
\end{center}
\end{figure}

The chemical evolution of protoplanetary disks has been investigated in detail by using robust chemical models (\cite{WKMH98,ah1999,wl00,Aea02,BCDH03,vZea03,IHMM04},\cite[Semenov et~al.~2004]{Red2},\cite{Willacy_ea06,IN_08a}). The current theoretical picture based on a steady-state prescription of the disk structure divides the disk into three zones, see Fig~\ref{scheme}. Before planets have formed and disk gas is dispersed, the dense midplane is well shielded from stellar and interstellar high-energy radiation. While its inner part can be heated up by accretion, the outer zone is cold, $T \sim 10-20$~K. The only ionization sources are cosmic ray particles and decay of short-living radionuclides, and thus matter remains almost neutral, with a low degree of turbulence. The molecular complexity in the midplane is determined by ion-molecule and surface reactions, with most molecules sitting on the grains. Adjacent to the midplane a warmer zone is located, which is partly shielded from stellar and interstellar UV/X-ray radiation. The complex cycling between efficiently formed gas-phase molecules, accretion onto dust surfaces, rapid surface reactions, and non-negligible desorption result in a rich chemistry (see for review \cite{Bergin_ea07}). The inner  part ($\lesssim 10-20$~AU) of this region in the disks around T Tauri stars can be substantially ionized by stellar X-ray radiation. The intermediate molecular layer is sufficiently dense ($\sim 10^5-10^6$~cm$^{-3}$) to excite most of the observed emission lines. Atop a hot and heavily irradiated surface layer is located, where C$^+$, light hydrocarbons, their ions, and other radicals like C$_2$H and CN are able to survive. This is the region where PAH and silicate emission features are produced.

\section{Theoretical constrains}
\begin{figure}
\includegraphics[width=0.45\textwidth]{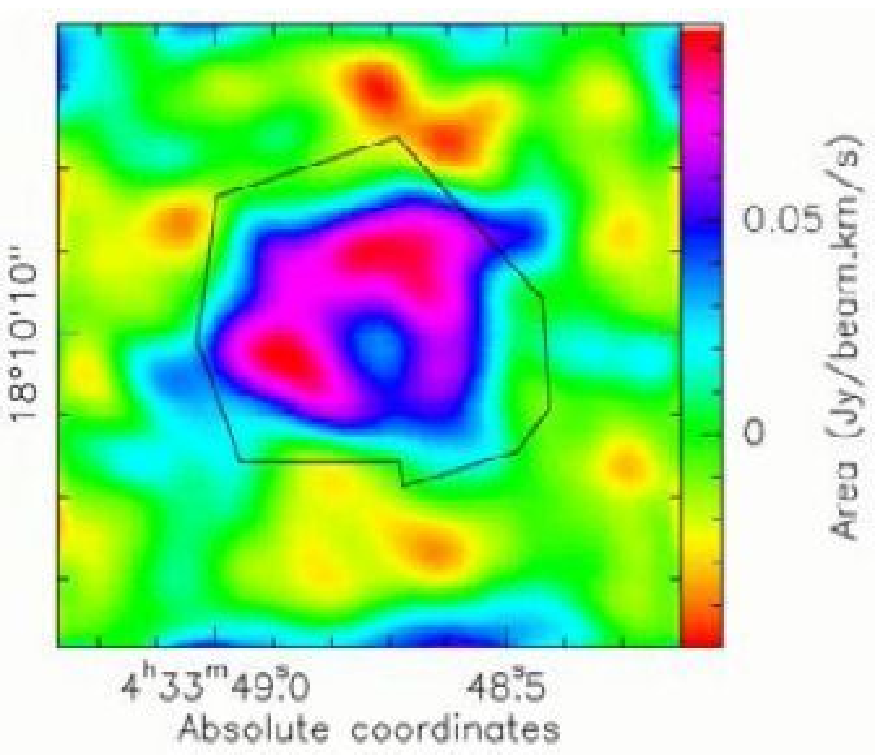}
\includegraphics[width=0.4\textwidth]{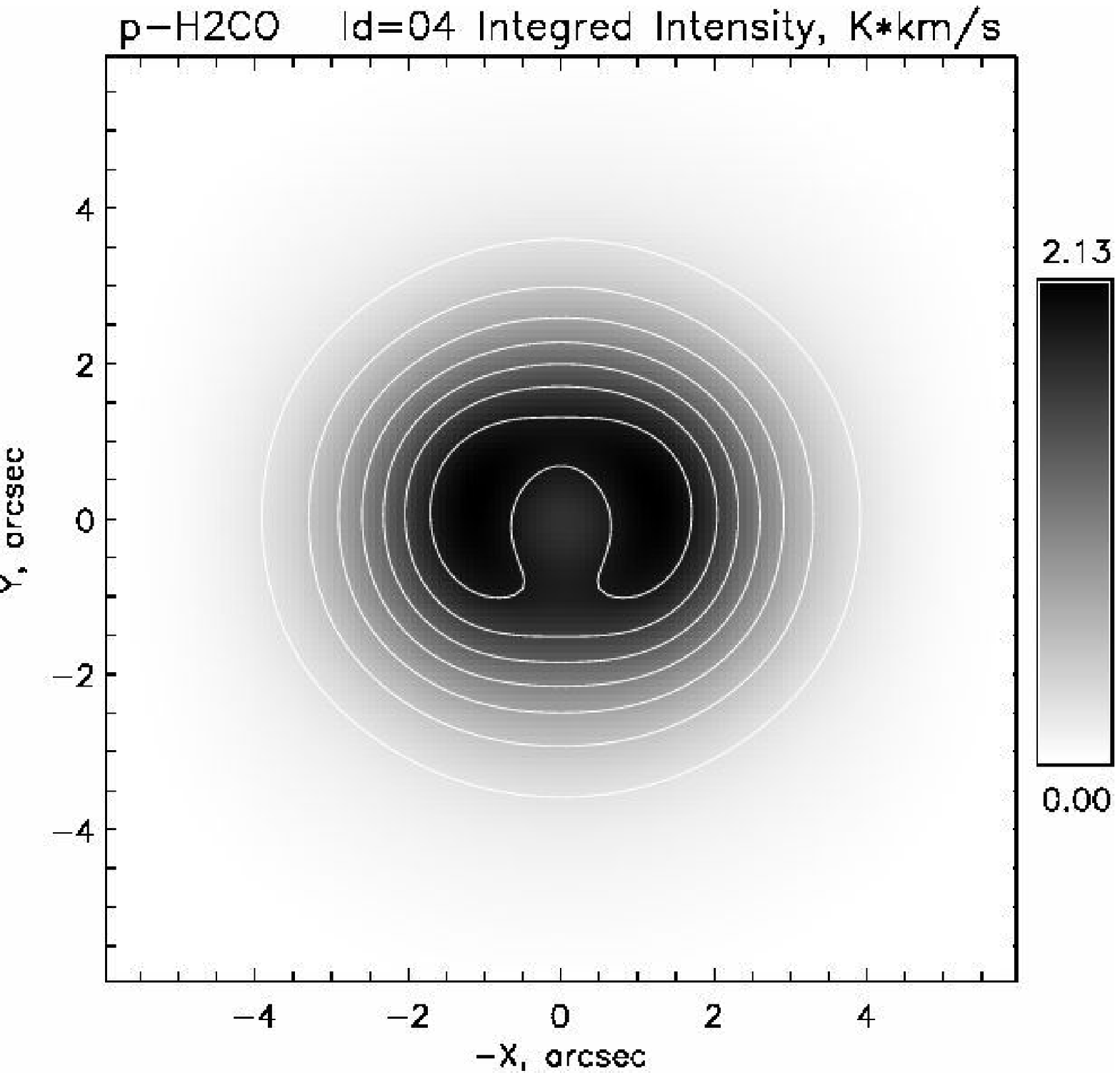}
\caption{(Left) The observed integrated intensity map of H$_2$CO\,(3--2) in the disk of DM Tau shows an asymmetric shell-like distribution with a chemical inner ``hole'' of $\sim 100$~AU in radius. (Right) The same features are present in the synthetic integrated intensity map of para-H$_2$CO\,(4--3) that is produced with a realistic disk physical and chemical model and line radiative transfer.}
\label{dm_tau_obs}
\end{figure}

Here we discuss a puzzling observation of a putative H$_2$CO inner cavity in the disk of DM~Tau. The Plateau de Bure interferometric image of the DM Tau disk at the 1.5$''$ resolution in the H$_2$CO (3$_{12}$-2$_{12}$) line is shown in Fig.~\ref{dm_tau_obs} (left panel). Despite high noise level, the H$_2$CO emission appears as asymmetric ring-like structure, with a dip in southern direction. To make a proper analysis of these data, a consistent combination of disk physical and chemical models along with radiative transfer in molecular lines is used. To simulate the disk physical structure we utilize a 1+1D flared disk model which is similar to the model of \cite{DAea99} with a vertical temperature gradient. The dust grains are modeled as compact amorphous silicate spheres of uniform 0.1~$\mu$m radius, with the opacity data taken from \cite{RP_opacities} and a dust-to-gas mass ratio of 100. The accretion rate is assumed to be $2\times 10^{-9}M_\odot$\,yr$^{-1}$, $\alpha=0.01$, and the disk outer radius is 8000~AU. We focus on the observable disk structure beyond the radius of $\sim 10$~AU. The total disk mass is 0.07\,$M_{\odot}$ and the disk age is 5~Myr (\cite[Pi\'etu et~al.~2007]{PDG_07}).

We assumed that the disk is illuminated by UV radiation from the central star with an intensity $\chi =410\,\chi _0$ at $100$~AU and by interstellar UV radiation with intensity $\chi _0$ in plane-parallel geometry (\cite{Dra78,vDea_06,Dutrey_ea07}). We model the attenuation of cosmic rays (CRP) by Eq.~(3) from \cite{Red2} with an initial value of the ionization rate $\zeta _\mathrm{CRP}=1.3\cdot 10^{-17}$~s$^{-1}$. In the disk interior ionization due to the decay of short-living radionuclides is taken into account, assuming an ionization rate of $6.5\cdot 10^{-19}$~s$^{-1}$ (\cite{FinocchiGail97}). The X-ray ionization rate in a given disk region is computed according to the results of \cite{zetaxa,zetaxb} with parameters for their high-metal depletion case and a total X-ray luminosity of $\approx 10^{30}$~erg\,cm$^{-2}$\,s$^{-1}$ (\cite{Glassgold_ea05}). The gas-phase reaction rates are taken from the RATE\,06 database (\cite{rate06}), while surface reactions together with desorption energies were adopted from the model of \cite{Garrod_ea06}. A standard rate approach to the surface chemistry modeling, but without H and H$_2$ tunneling was utilized (\cite{Katz_ea99}).

\begin{figure}
\includegraphics[width=0.3\textwidth,angle=90]{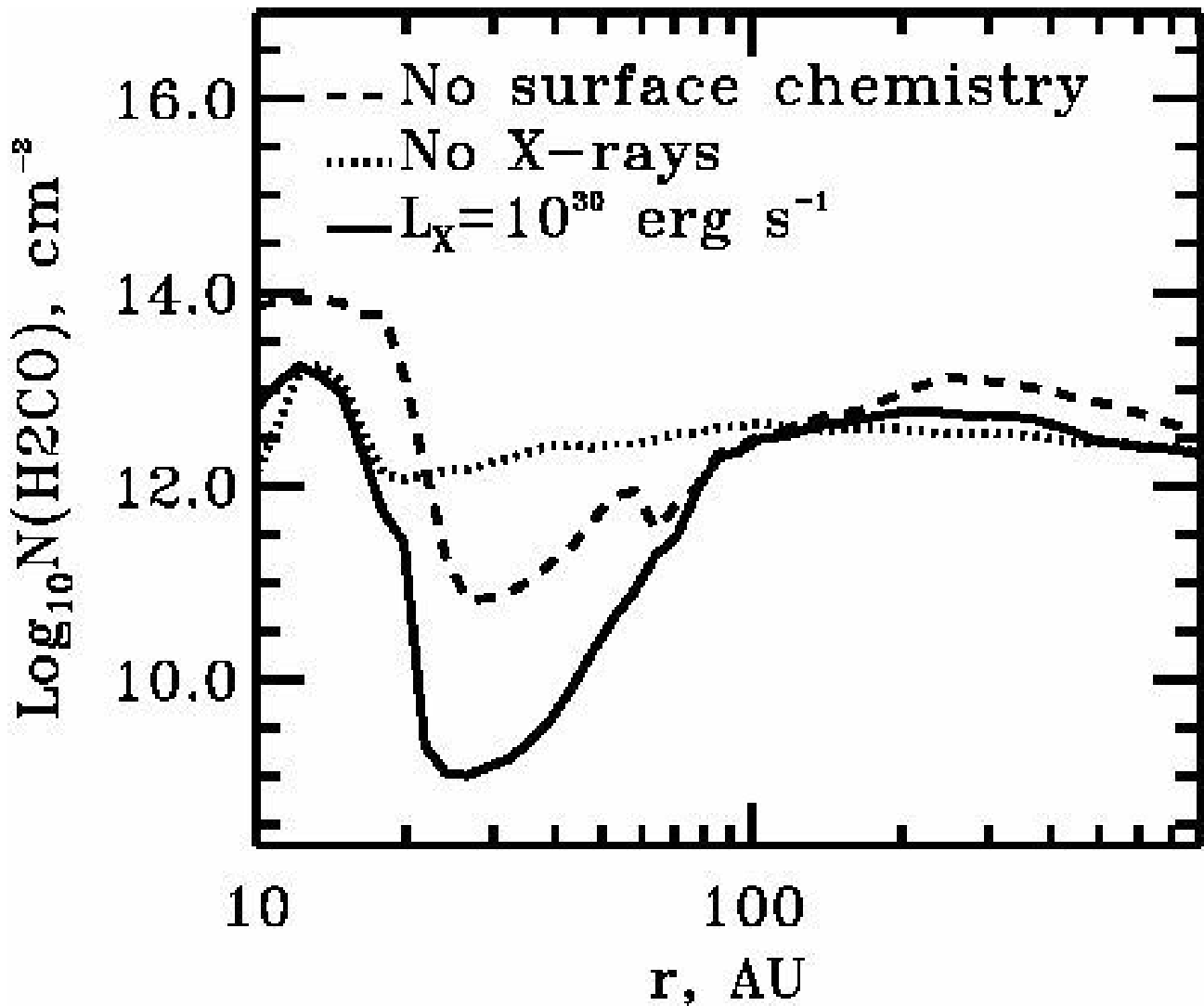}
\includegraphics[width=0.3\textwidth,angle=90]{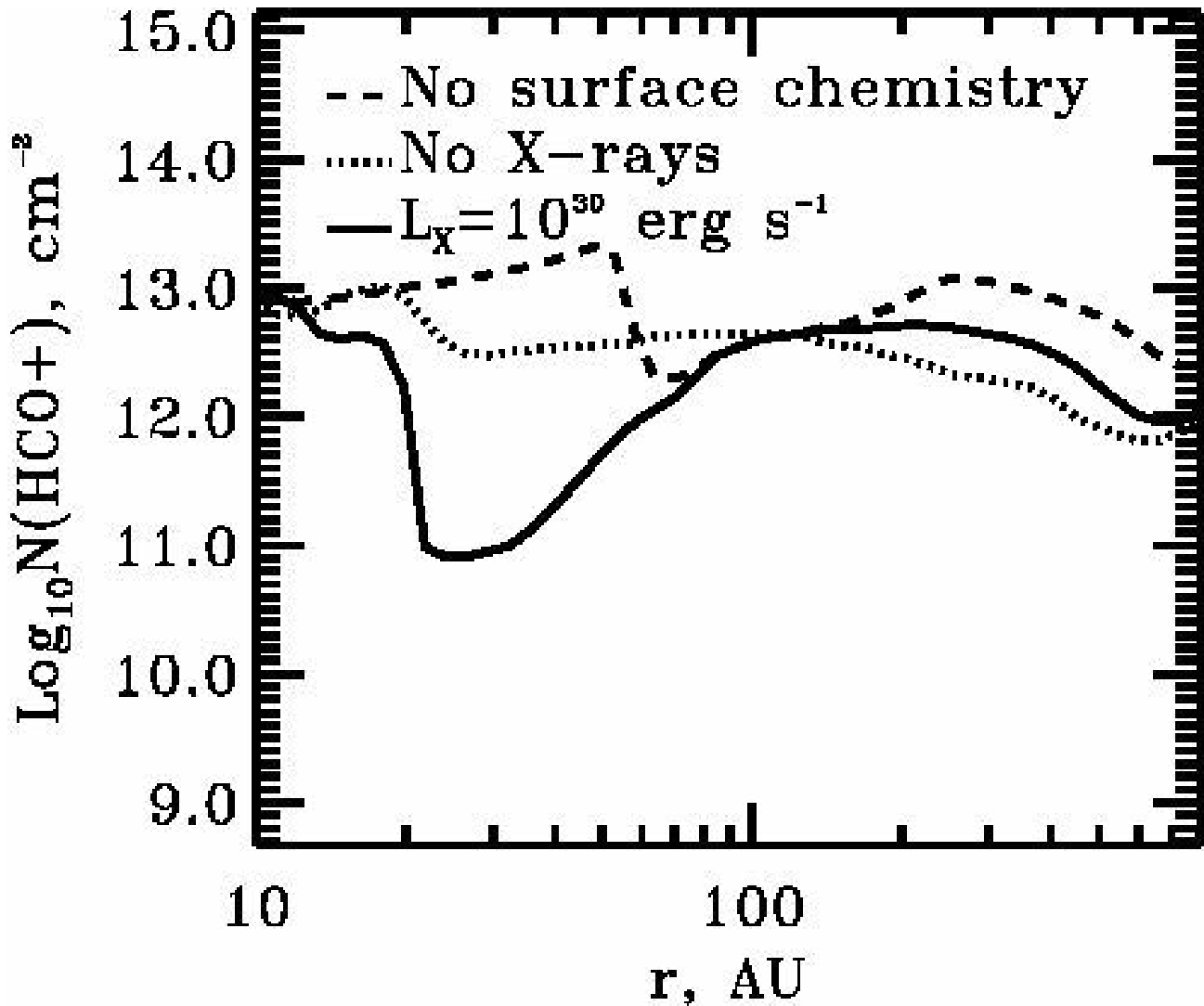}
\caption{(Left) The radial distribution of the HCO$^+$ column density in the DM Tau disk as computed with 3 different chemical models: 1) the stellar X-ray luminosity is assumed to be close to the observed value of $10^{30}$~erg/s (solid line), 2) no X-ray radiation penetrates into the inner disk (dotted line), and 3) the model without surface reactions but with $L_{\rm X}=10^{30}$~erg/s (dashed line). (Right) The same calculations but for the chemically related H$_2$CO molecule.}
\label{dm_tau_mod}
\end{figure}

Using the time-dependent chemical code ``ALCHEMIC''\footnote{\textit{www.mpia.de/homes/semenov}}, we simulated 5~Myr years of evolution in the DM~Tau disk, followed by 2D non-LTE line radiative transfer modeling with ``URANIA'' (\cite{Pavlyuchenkov_ea07}), see Fig.~\ref{dm_tau_mod}. The observed ring of H$_2$CO emission with a depression is fully reproduced by our model. We found two possible explanations why such a large-scale chemical hole can exist in the disk around DM Tau. The $\sim 100$~AU hole in the H$_2$CO emission is fully reproduced by both a disk model with X-ray driven chemical processes and somewhat less markedly in the model without surface chemistry. These two models predict different spatial distributions of molecular species, which can be tested by future interferometric observations.

Destruction of formaldehyde has important consequences for organic chemistry. The X-ray chemical model leads to the clearing of an inner hole of $\sim 100$~AU radius in all chemically related CO-bearing species, including HCO$^+$, by converting gas-phase CO into heavier CO$_2$-containing and chain-like hydrocarbon molecules. In contrast to CO, these heavier species are locked on dust surfaces in the inner disk region, where temperatures are lower than about 35-50~K (Fig.~\ref{dm_tau_mod}, solid line). This implies substantially different initial conditions with respect to the presence of complex organic molecules inside the planet-forming zone of protoplanetary disks, if this X-ray driven chemistry is important.

The less realistic model without surface chemistry shows the inner depression in column densities of highly saturated molecules only, like H$_2$O, NH$_3$, and to some extent H$_2$CO (see Fig.~\ref{dm_tau_mod}, dashed line). These species are formed on dust surfaces in a sequence of hydrogen addition reactions. Though at current stage we cannot fully distinguish between these two scenarios, for our understanding of the evolution of organic species in protoplanetary disks it will be of great importance to verify which of these explanations are valid.

\section{Summary}
We briefly overview recent progress in our understanding of chemical evolution in protoplanetary disks, from both the theoretical and observational perspective. A puzzling observation of the chemical inner hole visible in the spatial distribution of the H$_2$CO emission in the disk of DM~Tau is addressed theoretically. We found that such a hole can be explained either by the absence of efficient hydrogenation reaction on dust surfaces or efficient processing of disk matter by stellar X-ray radiation in the inner disk region, which was overlooked in previous studies. In future, when the Atacama Large Millimeter Array will become operational, the planet-forming zone of disks will be observable and this hypothesis can be verified. In general, chemo-dynamical models of disks together with interferometric observations well lead to a comprehensive understanding of the molecular inventory of protoplanetary disks.

\end{document}